\documentstyle[12pt,epsf,psfig]{article}
\begin{document}
\title{\bf  Lepton flavor violation two-body decays of quarkoniums}
\author { Wu-Jun Huo
\\
{\small\it
 Department of Physics, Peking University, Beijing $100871$, P. R.  China}
 \\
Tai-Fu Feng
\\{\small\it
 Institute of High Energy Physics, Chinese Academy of Sciences,
  P.O. Box $918(4)$},\\ {\small\it Beijing $100039$, P. R.
China }\\
Chong-Xing Yue\\
{\small\it
 College of Physics and Information Engineering, Henan Normal University,}
 \\{\small\it Xinxiang 453002, P. R. China}
}
\date{}
\maketitle
\begin{abstract}

In this paper
we firstly study various model-independent bounds on lepton flavor
violation (LFV)
in processes of $J/\Psi$, $\Psi'$ and $\Upsilon$ two-body decays, then
calculate their branch ratios
% By using the constraints from other ways, we  obtain
%the indirect bounds of ${\rm Br} (J/\Psi,\Psi',\Upsilon \to ll')$
in models of the leptoquark, $R$ violating MSSM and topcolor assisted
technicolor(TC2) models.

%respectively. It is shown that some of bounds would reach the
%experimental level and these new particles perhaps could  be seen
%or ruled out by the near future experiments.

\end{abstract}

 %\vspace*{0.5cm} \noindent
%PACS: 11.30.Er,12.60.-i,13.25.-Es,14.80.-j

\newpage

\section{Introduction}

At present, Standard model (SM)  not only has theoretical shortcoming but
 also must face to the experimental difficulties. The recent
 measurement of the muon anomalous magnetic moment by the experiment
 E821 \cite{e821} disagrees with the SM expectations
 %at more than 2$\sigma$ level.
 And there are also convincing evidences that neutrinos
 are massive and oscillate in flavor\cite{neutrino}. It seems to indicate the
  presence of new physics just round the corner will be in the leptonic part.
As probing new physics, lepton flavor violation (LFV) processes
 have as natural consequence an increased interest experimentally and
 theoretically. Several experiments that may considerably improve on lepton
 flavor violating processes, such as the two-body LFV of bosons, $\mu$ and
$\tau$ LFV decays, are under consideration. For example, the TESLA
Linear Collider project will work at the $Z$ resonance, reaching a
$Z$ production rate of $10^9$ per year \cite{zrate}. A number of
theoretical studies are devoted to these LFV decays to investigate
the new physics effects \cite{number,number2, zhang1,yue}.  For
$J/\Psi$ meson, BES \cite{bes} has accumulated about $10^7$ -
$10^8$ $J/\Psi$ events which can be available to probe lepton
flavor violation of $J/\Psi$ \cite{zhang2}.   In this note, we
study the LFV processed of the heavy vector bosons, $J/\Psi$,
$\Psi'$ and $\Upsilon$ with some models beyond SM.

  The LFV decays of $J/\Psi$, $\Psi'$ and
$\Upsilon$, such as $J/\Psi, \Psi',\Upsilon \to \mu e, \tau l$,
are absent in SM at tree level because of the strong GIM
suppression. They are strongly suppressed by powers of small
neutrino masses and have very small branching ratios.
Experimentally, there are no clear bounds of these decays. Then,
such decays give therefore existence room of new physics. Ref.
\cite{zhang2} discuss these flavor decays by using the simple
"unitarity inspired" relations and get rather strong
model-independent constraints on these two-body LFV processes
\begin{eqnarray}
{\rm Br}(J/\Psi \to\mu e)&\leq& 4\times 10^{-13}\\
{\rm Br}(J/\Psi \to\tau {\bar l})&\leq& 6\times 10^{-7}\\
{\rm Br}(\Upsilon \to\mu e)&\leq& 2\times 10^{-9}\\
{\rm Br}(\Upsilon \to\tau {\bar l})&\leq&  10^{-2}
\end{eqnarray}
From the above constraints, we can see perhaps BES  could observe some of
these processes.

   In fact, in many models beyond SM, there exist many bosons
(scalars or vectors), such as leptoquarks in GUTs, sleptons in
SUSY  and $Z'$ in technicolor models, which can induce the LFV
decays at tree level and contribute  large branching ratios. In
many GUTs, the existence of leptoquarks is predicted, which has
been actively searched in many collider experiments \cite{lepexp}.
These new particles, which carry both the lepton and quark
numbers, can couple to a current comprising of a lepton and a
quark \cite{he,hewett}. Thus, they can lead to the vertices
$J/\Psi\mu e$, $J/\Psi\mu\tau$, and so on. Recently, Refs.
\cite{cheung,datta} investigate the muon anomalous magnetic moment
with the leptoquarks and get the restricting parameter space.
Leptoquarks can induce the quarknioums decay to $ll'$ through
$t$-channel.

  The similar situation is in SUSY models without lepton number
  conservation and R-parity which the squarks play the same role as
leptoquarks in GUTs\cite{mssm,rparity,susy1,susy2}. Although the
present experiment constraint those couplings in various ways, the
R-parity violating coupling may give large contribution to
$J/\Psi\mu e$, $J/\Psi\mu\tau$, and so on.

  In topcolor assisted technicolor (TC2) models\cite{tc2}, when the
   non-universal interactions, topcolor interactions are written in the
   mass eigenstates, it may lead to the flavor changing coupling vertices
   of the new gauge boson $Z'$, such as $Z'\mu e$, $Z'\mu\tau$. Thus, the
   $Z'$ has significant contributions to the lepton flavor changing process, like $\mu\to
3e$, and may have severe bound on the mass $M_{Z'}$ \cite{yue}.
$Z'$ can also, but different from leptoquarks and sleptons, induce
the decays of quarknioums to $ll'$ through $s$-channel.

  We  investigate the bounds of $J/\Psi,\Psi',\Upsilon
  \to ll'$ with model-independent analysis in sec. 2. Then, in sec.3, we
  study these bounds with model-dependent analysis in three models,
  leptoquarks, SUSY and technicolor,  respectively. We give our
  conclusion in the last section.

\section{Bounds of  $V_i \to ll'$ with  model independent analysis}

Considering that a vector boson $V_i$, such as $J/\Psi$, $\Psi'$
and $\Upsilon$, couples to ${\bar l}l'$, the effective coupling
between the vector boson $V_i$ and ${\bar l}l'$ as
\begin{eqnarray}
{\cal L}_{\rm eff} = g'_{Vll'} {\bar l} \Gamma_\alpha l' V^\alpha +h.c.
\end{eqnarray}
and \cite{number}
\begin{eqnarray}
\Gamma_\alpha =\left( \gamma_\alpha A^L_1
+i\sigma_{\alpha\beta}\frac{q^\beta}{M}A^L_2
+\frac{q_\beta}{M}A^L_3 \right)P_L +(L\leftrightarrow R),
\end{eqnarray}
where mass scale $M$ is introduced to make the form factors
$A^{L,R}_{2,3}$ dimensionless. For on-shell vector mesons, the
$A^{L,R}_3$ form factors do not contribute. With
this langragian, we can calculate the decay $\tau\to ll'{\bar l}'$
(see Fig. 1). In the limit $m_e \to 0$, we can obtain
\begin{eqnarray}
\Gamma (\tau\to ll'{\bar l}' ) &=&\frac{{g'}^2_{V \tau l} g^2_{V
l'l'}m^5_\tau Q^2 \alpha^2}{192\pi^3 M^8_V} \left(|A^L_1 -\frac{m_\tau
A^R_2}{2M}|^2 + |A^R_1 -\frac{m_\tau A^L_2}{2M}|^2
+\frac{3}{20}|\frac{m_\tau A^L_2}{M}|^2 \right)\nonumber\\
&&+(L\leftrightarrow R),
\end{eqnarray}
where $Q$ is the charge of quark in the quarkoniums, $Q=2/3$ for
$J/\Psi$ and $\Psi'$ and $Q=-1/3$ for $\Upsilon$.  $g_{Vl'l'}$ is
the vector meson decay constants and no relate to the effective
coupling ${g'}_{V\tau l}$.

 Similarly, we can get LFV decay width
\begin{eqnarray}
\Gamma ( V\to \tau l) =\frac{{g'}^2_{V \tau l}}{24 \pi}
\frac{m^2_\tau}{M_V}
(1+2\frac{M^2_V}{m^2_\tau})(1-\frac{m^2_\tau}{M^2_V})^2 \left( |A^L_1 |^2 +
\frac{1}{2}|\frac{m_V A^L_2}{M}|^2 +(L\leftrightarrow R) \right).
\end{eqnarray}

Using above equations and compared to the standard decays $\tau
\to \nu_\tau l \nu_l$
\begin{eqnarray}
\Gamma (\tau\to \nu_\tau l \nu_l ) =\frac{G^2_F m^5_\tau}{192\pi^3}
\end{eqnarray}
and
\begin{eqnarray}
\Gamma ( V\to l' l') =\frac{4\pi\alpha^2
Q^2}{3}\frac{g^2_{Vl'l'}}{M^3_V},
\end{eqnarray}
we obtain
\begin{eqnarray}
\frac{{\rm Br}(\tau\to l l'{\bar l}')}{{\rm Br}(V\to \tau l)\cdot
{\rm Br}(V\to l' {\bar l}')}&=&\frac{\Gamma (\tau\to l l'{\bar
l}')}{\Gamma (\tau\to \nu_\tau l'{\bar \nu_{l'}})}\cdot
\frac{\Gamma^2_V}{\Gamma (V\to \tau l)\cdot \Gamma (V\to l' {\bar
l}')}\nonumber\\ &&\times {\rm Br}(\tau\to \nu_\tau l'{\bar
\nu_{l'}}) \nonumber\\ &=&\frac{18\Gamma^2_V \cdot {\rm
Br}(\tau\to \nu_\tau l'{\bar \nu_{l'}})}{G^2_F m^2_\tau M^4_V
(1+2\frac{M^2_V}{m^2_\tau})(1-\frac{m^2_\tau}{M^2_V})^2} \times
{\cal A},
\end{eqnarray}
where $\Gamma_V$ is the total decay width of the vector bosons $V$ and
\begin{eqnarray}
{\cal A}=\frac{(|A^L_1 -\frac{m_\tau A^R_2}{2M}|^2 +
|A^R_1 -\frac{m_\tau A^L_2}{2M}|^2
+\frac{3}{20}|\frac{m_\tau A^L_2}{M}|^2 )+(L\leftrightarrow R)}
{(|A^L_1 |^2 +\frac{1}{2}|\frac{m_V A^L_2}{M}|^2
+(L\leftrightarrow R))}.
\end{eqnarray}

  By using the experimental bounds of ${\rm Br}(\tau\to ll'{\bar l}') $ and the experimental
values of ${\rm Br}(V\to l'l')$ and ${\rm Br}(\tau\to \nu_\tau
l'{\bar \nu_{l'}})$ \cite{PDG}, we could obtain the bounds of
$V\to \tau l$.
 For discussing the bounds, as like doing
in \cite{number}, we consider two limiting cases:
\begin{itemize}
\item When $|A^L_1|$ or $|A^R_1| \gg \frac{m^2_\tau}{M^2}|A^{L,R}_2|, {\cal
A}\approx 1$. From eq.(7), we get
\begin{eqnarray}
{\rm Br}(J/\Psi\to \tau l) &<& 1.0\times 10^{-7}\\
{\rm Br}(\Psi' \to\tau {\bar l})&\leq& 0.7\times 10^{-7}\\
 {\rm
Br}(\Upsilon\to \tau l) &<&8.0\times 10^{-5}
\end{eqnarray}
Similarly, by using ${\rm Br}(\mu\to e^- e^+ e^-) \leq 10^{-12}$
and the experimental values of ${\rm Br}(V\to e^- e^+)$ and ${\rm
Br}(\mu\to \nu_\mu e {\bar \nu_{e}})$ \cite{PDG}
\begin{eqnarray}
{\rm Br}(J/\Psi\to \mu e) &<& 2\times 10^{-13}\\
{\rm Br}(\Psi' \to\mu e)&\leq& 1.2\times 10^{-13}\\
{\rm Br}(\Upsilon\to \mu e) &<&1.7\times 10^{-9}
\end{eqnarray}
\item When $|A^L_1|$ or $|A^R_1| \ll \frac{m^2_\tau}{M^2}|A^{L,R}_2|, {\cal
A}\approx \frac{13}{10}\frac{m^2_\tau}{M^2_V}$. We get
\begin{eqnarray}
{\rm Br}(J/\Psi\to \tau l) &<& 3.6\times 10^{-7}\\
{\rm Br}(\Psi' \to\tau {\bar l})&\leq& 2.5\times 10^{-7}\\
{\rm Br}(\Upsilon\to \tau l) &<&2.9\times 10^{-4}
\end{eqnarray}
Similarly,
\begin{eqnarray}
{\rm Br}(J/\Psi\to \mu e) &<& 5.3\times 10^{-13}\\
{\rm Br}(\Psi' \to\mu e)&\leq& 3.6\times 10^{-13}\\
{\rm Br}(\Upsilon\to \mu e) &<&1.5\times 10^{-8}
\end{eqnarray}
\end{itemize}
From the above equations, we find some of them ( Eqs. (13) and
(19)) can reach the current experimental level of BES.

  \section{Bounds of  $J/\Psi \to ll'$
with  model-dependent analysis}

\subsection{In leptoquark model}

Many models beyond SM, like GUTs, naturally contain leptoquarks
which  can couple to a lepton-quark pair. This can induce the LFV
two-body decays of $J/\Psi$ and $\Upsilon$ through $t$-channel
(see Fig. 2).

The Leptoquarks contributing to these diagrams are $\Phi_1$ and $\Phi_3$
\cite{datta}. Their couplings are
\begin{eqnarray*}
\Phi_1 &: &\,\,\,\,\left[\lambda_{ij}^{(1)} {\bar Q}_{L j} e_{R i}
 + {\tilde \lambda}_{ij}^{(1)} {\bar u}_{R j} L_{Li}\right]\Phi_1,
  \nonumber\\ \Phi_3&:&\,\,\,\,\left[\lambda_{ij}^{(3)} {\bar Q}_{L j}^c
   L_{L i}+ {\tilde \lambda}_{ij}^{(3)} {\bar u}_{R j}^c
   e_{Ri}\right]\Phi_3.
\end{eqnarray*}
Confining ourselves to terms involving the $\mu$ and $\tau$,
$J/\Psi$ and $\Phi_1$, the relevant part of the interaction
Lagrangian can be parametrized as
\begin{eqnarray}
{\cal L}^{Leptoquark}_{eff}
& =&{\bar c}(\lambda^A_L P_L +\lambda^A_R P_R
)\mu \Phi_A  \nonumber\\
& &+{\bar c}(\lambda^A_L P_L +\lambda^A_R P_R
)\tau \Phi_A + h.c.
\end{eqnarray}
where $\Phi$ is one of the above two leptoquarks, $\lambda_{L,R}$
is the structure of the chiral couplings and $P_{L,R} =(1\mp
\gamma^5 )/2$.

The decay width is
\begin{eqnarray}
\Gamma (J/\Psi\to\mu\tau) &=& \frac{\bf |p|}{32\pi^2 M_{J/\Psi}}
\int |{\cal M}|^2 d\Omega \nonumber\\
&=&\frac{g^2_{J/\Psi}}{96\pi}\frac{m^2_\tau}{M_{J/\Psi}}(1+2
\frac{M^2_{J/\Psi}}{m^2_\tau})(1-
\frac{m^2_\tau}{M^2_{J/\Psi}})^2\cdot \frac{|\lambda^{c\mu}_L
\lambda^{c\tau}_L |^2 + |\lambda^{c\mu}_R \lambda^{c\tau}_R |^2
}{M^4_\Phi},
\end{eqnarray}
where $g_{J/\Psi}$ is ${J/\Psi}$ decay constant.  Compared to
electromagnetic decay $J/\Psi \to e^+ e^- $ through $\gamma$,
\begin{eqnarray}
\Gamma (J/\Psi\to e^+ e^- )=\frac{16\pi}{27}\alpha^2
\frac{g^2_{J/\Psi}}{M^3_{J/\Psi}}, \end{eqnarray}
we get
\begin{eqnarray}
{\rm Br} (J/\Psi\to\mu\tau) &=&\frac{9}{2^9 \pi^2 \alpha^2 }
m^2_\tau M^2_{J/\Psi} (1+2\frac{M^2_{J/\Psi}} {m^2_\tau})(1-
\frac{m^2_\tau}{M^2_{J/\Psi}})^2 \cdot \frac{|\lambda^
{c\mu}_L\lambda^{c\tau}_L |^2 + \lambda^{c\mu}_R \lambda^{c\tau}_R
|^2 }{M^4_\Phi}  \nonumber \\ & &\times {\rm Br} (J/\Psi\to e^+
e^- )
\end{eqnarray}
We take the  experimental value, $({\rm
Br} (J/\Psi\to e^+ e^- )=(6.02\mp 0.19 )\%$ \cite{PDG}.
To get the constraints of
$(|\lambda^{c\mu}_L \lambda^{c\tau}_L |^2 + |\lambda^{c\mu}_R
\lambda^{c\tau}_R |^2 )/M^4_\Phi$, we consider another lepton flavor decay
$\tau\to\mu\gamma$ through leptoquarks (see Fig. 3).
Compared to electroweak decay $\tau\to \mu \nu_\tau {\bar
\nu_\mu}$, this gives a branching ratio of
\begin{eqnarray}
{\rm Br} (\tau\to\mu\gamma)&=&\frac{3}{2^9 \pi^2 G^2_F}\cdot
\frac{|\lambda^{c\mu}_L \lambda^{c\tau}_L |^2 + \lambda^{c\mu}_R
\lambda^{c\tau}_R |^2 }{M^4_\Phi} \nonumber \\ & &\times  {\rm Br}
(\tau\to\mu\nu_\tau {\bar \nu_\mu}),
\end{eqnarray}
where $G_F$ is the effective electroweak couplings. By using the
experimental values, $ {\rm Br} (\tau\to\mu\gamma) < 1.1 \times
10^{-6}$ and  ${\rm Br} (\tau\to\mu\nu_\tau {\bar \nu_\mu})=
(17.37\pm 0.09 )\times 10^{-6}$, we can obtain
\begin{equation}
\frac{|\lambda^{c\mu}_L \lambda^{c\tau}_L |^2 + |\lambda^{c\mu}_R
\lambda^{c\tau}_R |^2 }{M^4_\Phi} <1.5\times 10^{-10}
\end{equation}
Then, we can obtain the bound of ${\rm Br} (J/\Psi\to\mu\tau)$
with a scalar leptoquark is
\begin{equation}
{\rm Br} (J/\Psi\to\mu\tau) < 3.0\times 10^{-8}.
\end{equation}
Similarly, we get
\begin{eqnarray}
{\rm Br} (J/\Psi\to\mu e) &<&3.5\times 10^{-15},\\
{\rm Br} (\Psi'\to\mu\tau) &<& 9.3\times 10^{-9},\\
{\rm Br} (\Psi'\to\mu e) &<&1.1\times 10^{-15},\\
 {\rm Br}(\Upsilon\to\mu\tau) &<&1.3\times 10^{-7},\\
 {\rm Br}(\Upsilon\to\mu e) &<&1.6\times 10^{-14}.
\end{eqnarray}

\subsection{In SUSY model}

In the supersymmetry model without R-parity and lepton number
conservation, rare processes $J/\Psi \Upsilon \to ll'$ is
induced by squark through $t$-channel, (see Fig. 2).

   The superpotential for the lepton number violation supersymmetry
   can be written as:
\begin{eqnarray}
&&{\cal W}={\cal W}_{_{\rm MSSM}}+{\cal W}_{_{/\!\!\! L}}\;.
\label{superpotential}
\end{eqnarray}
The ${\cal W}_{_{\rm MSSM}}$ represents the R-parity conservation sector
supersymmetry and can be found in literatures\cite{mssm}. The R-parity
violation sector superpotential is
\begin{eqnarray}
{\cal W}_{_{/\!\!\! L}}=\epsilon_{ij}\lambda_{_{\rm IJK}}\hat{L}_i^{^I}
\hat{L}_j^{^J}\hat{R}^{^K}+\epsilon_{ij}\lambda_{_{\rm
IJK}}^\prime\hat{L}_i^{^I} \hat{Q}_j^{^J}\hat{D}^{^K}
\label{rparity}
\end{eqnarray}
with $\hat{L}^{^I}$ represents the $I$-th generation lepton
superfields and $\hat{Q}^{^I}$ represents the $I$-th generation
quark superfields, which are all the doublet of $SU(2)$ group. The
$\hat{R}^{^I}\;,\hat{D}^{^I}$ are the $I$-th generation $SU(2)$
singlet lepton- and quark- superfields. Here, we have ignore the
bilinear lepton number violation terms in the
superpotential\cite{rparity}. Although there are 36 trilinear
R-parity couplings in the superpotential Eq.(\ref{rparity}), our
computation involve only two trilinear couplings
$\lambda_{_{222}}^\prime$ and $\lambda_{_{223}}^\prime$.

The calculation is similar to that
of leptoquarks. The relative effective Lagrangian can be written as
\begin{equation} {\cal L}^{SUSY}_{eff}
=\frac{\lambda_{_{222}}^\prime\lambda_{_{223}}^\prime
}{4M^2_\Phi}({\bar c} P_L \mu {\bar \tau } P_R c+{\bar c} P_L \tau
{\bar \mu } P_R c),
\end{equation}
where $M_{\Phi}^2$ is the mass of squark.

From this Lagrangian, we can get decay width with squark $\Phi$ is
\begin{eqnarray}
\Gamma (J/\Psi\to\mu\tau) = \left(
\frac{(\lambda_{_{222}}^\prime\lambda_{_{223}}^\prime)^2
g^2}{3\times 16^2 \pi M^4_{\Phi}}\right)
\frac{m^2_\tau}{M_{J/\Psi}}(1+2\frac{M^2_{J/\Psi}} {m^2_\tau})
(1-\frac{m^2_\tau}{M^2_{J/\Psi}})^2.
\end{eqnarray}
Comparing to ordinary decay $J/\Psi \to e^+ e^- $ (Eq. (8)), we
obtain
\begin{eqnarray}
{\rm Br} (J/\Psi\to\mu\tau) &=&\frac{9}{8\times 16^2 \pi^2 \alpha^2}
\left(\frac{(\lambda_{_{222}}^\prime\lambda_{_{223}}^\prime)^2
} { M^4_{\Phi}}\right) m^2_\tau M^2_{J/\Psi}(1+2\frac{M^2_{J/\Psi}}
{m^2_\tau})(1-\frac{m^2_\tau}{M^2_{J/\Psi}})^2 \nonumber\\ & &\times
{\rm Br} (J/\Psi\to e^+ e^- )
\end{eqnarray}

  As for tribilinear coupling constants, we adopt the single coupling
hypothesis, where a single coupling constant is assumed to dominate over all
the others, so that each of the coupling constants contributes once at a
time\cite{susy1}. The analysis of tree level $/\!\!\! R_{_{\rm P}}$
contributions to the D-mesons three body decays, $D\rightarrow
K+l+\nu\,;D\rightarrow K^*+l+\nu$ yields the bounds\cite{susy2}
$$\lambda_{_{22K}}^\prime<0.18,\; \;K=1,\;2,\;3.$$
If the supersymmetry is weak-scale theory with $M_{\Phi}=100{\rm GeV}$,
 we can obtain the bound of ${\rm Br} (J/\Psi\to\mu\tau)$
with sleptons is
\begin{equation}
{\rm Br} (J/\Psi\to\mu\tau) < 5.0\times 10^{-9}.
\end{equation}
Similarly, we get
\begin{eqnarray}
{\rm Br} (J/\Psi\to\mu e) &<&5.7\times 10^{-16},\\
{\rm Br} (\Psi'\to\mu\tau) &<& 1.8\times 10^{-9},\\
{\rm Br} (\Psi'\to\mu e) &<&1.9\times 10^{-16},\\
 {\rm Br}(\Upsilon\to\mu\tau) &<&2.2\times 10^{-8},\\
  {\rm Br}(\Upsilon\to\mu e) &<& 2.5\times 10^{-15}.
\end{eqnarray}

\subsection{In TC2 models}

To solve the phenomenological difficulties of traditional
technicolor theory, TC2 theory\cite{tc2} was proposed by combing
technicolor interactions with topcolor interactions for the third
generation at the energy scale of about 1 TeV. TC2 models predict
the existence of the new gauge boson $Z'$, which lead to the LFV
coupling vertices, $Z' ll'$\cite{yue}. Thus, it can give
significant contributions to some LFV processes. In TC2 models,
the contributions of $ Z' $ to the LFV process $J/\Psi \to
\mu\tau$ can be induced through $s$-channel, (see Fig. 4).

  The couplings of the new gauge boson $ Z' $ to ordinary
  fermions, which are related to the LFV process $J/\Psi \to
\mu\tau$, can be written as:
\begin{equation}
{\cal L}^{Z'}_{eff} =-\frac{g_1 \tan\theta'}{6}Z'_{\mu}[{\bar c_L}
\gamma^\mu c_L + 2{\bar c_R} \gamma^\mu c_R +....]-
\frac{g_1}{2}Z'_{\mu}[k_{\tau\mu} ({\bar \mu_L} \gamma^\mu \tau_L
 + 2{\bar \mu_R} \gamma^\mu \tau_R )+ .....],
\end{equation}
where $g_1$ is the $U(1)_y$ coupling constant at the scale $
\Lambda_{TC} $, $k_{\tau\mu}$ is the flavor mixing factor, and
$\theta'$ is the mixing angle. With the above Lagrangian, we can
obtain the decay width contributed by the new gauge boson $ Z' $:
\begin{eqnarray}
\Gamma (J/\Psi\to\mu\tau) = \left( \frac{\pi k_1 \tan^4 \theta'
}{12M^2_{Z'}}\right)^2 \frac{g^2 (k^2_L +4k^2_R )m^2_\tau}{12\pi M_{J/\Psi}}
(1+2\frac{M^2_{J/\Psi}}
{m^2_\tau})(1-\frac{m^2_\tau}{M^2_{J/\Psi}})^2
\end{eqnarray}
Compared  to ordinary decay $J/\Psi \to e^+ e^-$, we obtain
\begin{eqnarray}
{\rm Br} (J/\Psi\to\mu\tau) &=&\frac{9}{32\times 12^2 \alpha^2}
\left( \frac{ k_1 \tan^4 \theta' }{M^2_{Z'}}\right)^2 (k^2_L
+4k^2_R )m^2_\tau M^2_{J/\Psi} (1+2\frac{M^2_{J/\Psi}}{m^2_\tau})\nonumber\\
& & \times(1-\frac{m^2_\tau}{M^2_{J/\Psi}})^2  {\rm Br} (J/\Psi\to
e^+ e^- )
\end{eqnarray}
  Using the results of Ref.\cite{yue}, we can obtain the bound of ${
  \rm Br} (J/\Psi\to\mu\tau)$ with $Z'$ is
\begin{equation}
{\rm Br} (J/\Psi\to\mu\tau) < 3.3 \times 10^{-8}.
\end{equation}
Similarly, we get
\begin{eqnarray}
{\rm Br} (J/\Psi\to\mu e) &<&3.8 \times 10^{-15},\\
{\rm Br} (\Psi'\to\mu\tau) &<& 1.0\times 10^{-8},\\
{\rm Br} (\Psi'\to\mu e) &<&1.2\times 10^{-15},\\
 {\rm Br}(\Upsilon\to\mu\tau) &<&1.3 \times 10^{-7},\\
 {\rm Br}(\Upsilon\to\mu e) &<&1.6\times 10^{-14}.
\end{eqnarray}

 \section{Conclusions}

we have investigated the bounds of lepton flavor violation
processes of $J/\Psi$ and $\Upsilon$ two-body decays in
leptoquarks, SUSY and TC2 models, respectively. We used the
constraints of couplings from other ways to obtain the indirect
bounds of ${\rm Br} (J/\Psi,\Upsilon \to ll')$  with model
independent. It is interesting  that some results would get th
experimental level. And it is shown that these new particles
perhaps can be seen or ruled out by the near future experiments.

\section*{Acknowledgments}
We thank Professor Xinmin Zhang for suggestions on this project. One of
the authors (W. J. Huo) acknowledges  supports from
the Chinese Postdoctoral  Science Foundation and CAS K.C. Wong Postdoctoral
Research Award  Fund.  The work of C. X. Yue was supported by the National
Natural SScience Foundation of China(I9905004), the Excellent Youth
Foundation of Henan Scientific Committee(9911), and Foundation of
Henan Educational Committee.

\newpage
\begin{figure}
\vskip 0cm \epsfxsize=20cm \epsfysize=20cm \centerline{
\epsffile{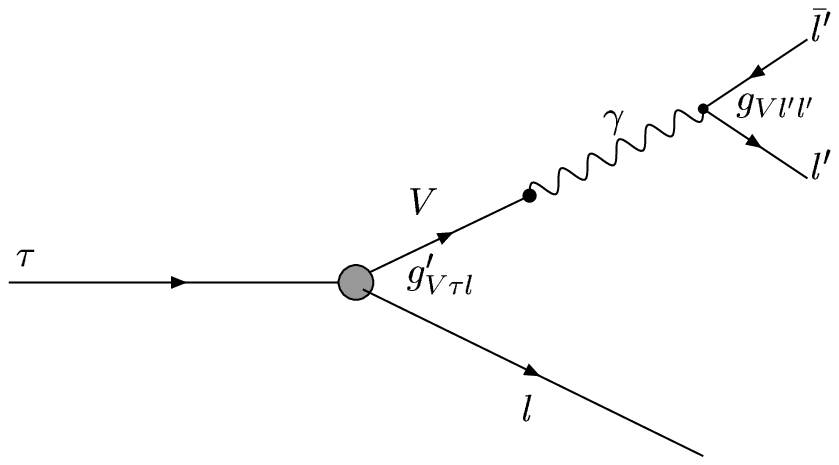}} \vskip -14cm \caption{Diagram of LFV decays
$\tau\to l{\bar l'}l'$ through vector mesons, $J/\Psi$,
$\Upsilon$}
\end{figure}

\newpage

\begin{figure}
\vskip 0cm \epsfxsize=20cm \epsfysize=20cm \centerline{
\epsffile{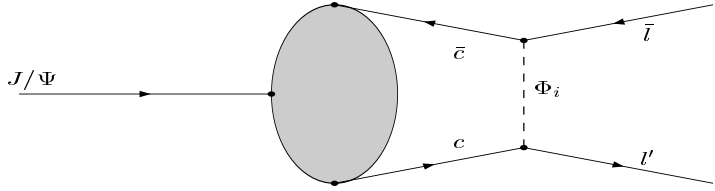}} \vskip -14cm \caption{Diagram
$J/\Psi\to\mu\tau$ through letoquarks or sleptons.}
\end{figure}

\newpage
\begin{figure}
\vskip 0cm \epsfxsize=20cm \epsfysize=20cm \centerline{
\epsffile{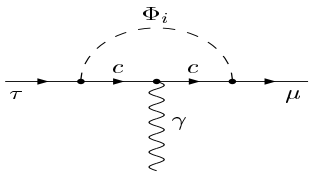}} \vskip -14.0cm \caption{Diagram of $\tau\to
\mu\gamma$ through letoquarks.}
\end{figure}
\newpage

\begin{figure}
\vskip 0cm \epsfxsize=20cm \epsfysize=20cm \centerline{
\epsffile{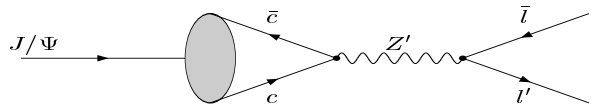}} \vskip -14cm \caption{Diagram of $J/\Psi
\to\mu\tau$ in TC2 models.}
\end{figure}

\end{document}